\documentclass[sigconf]{acmart}
\AtBeginDocument{%
  \providecommand\BibTeX{{%
    \normalfont B\kern-0.5em{\scshape i\kern-0.25em b}\kern-0.8em\TeX}}}

\copyrightyear{2022} 
\acmYear{2022} 
\setcopyright{rightsretained} 

\acmConference[ICPP '22]{51st International Conference on Parallel Processing}{August 29-September 1, 2022}{Bordeaux, France}
\acmBooktitle{51st International Conference on Parallel Processing (ICPP '22), August 29-September 1, 2022, Bordeaux, France}\acmDOI{10.1145/3545008.3545072}
\acmISBN{978-1-4503-9733-9/22/08}

\acmSubmissionID{64}

\usepackage[colorinlistoftodos]{todonotes}

\newcommand{\fix}[1]{{\color{black} #1}}

\usepackage{multirow}
\usepackage{makecell} % for more vertical space in cells
\setcellgapes{5pt}

\begin{document}

\title{FLOPs as a Discriminant for Dense Linear Algebra Algorithms}

\author{Francisco López}
\affiliation{%
  \institution{Umeå Universitet}
  \city{Umeå}
  \country{Sweden}}
\email{flopz@cs.umu.se}

\author{Lars Karlsson}
\affiliation{%
  \institution{Umeå Universitet}
  \city{Umeå}
  \country{Sweden}}
\email{larsk@cs.umu.se}

\author{Paolo Bientinesi}
\affiliation{%
 \institution{Umeå Universitet}
 \city{Umeå}
 \country{Sweden}}
\email{pauldj@cs.umu.se}

\renewcommand{\shortauthors}{López, Karlsson, and Bientinesi}

\begin{abstract}

Expressions that involve matrices and vectors, known as linear algebra expressions, are commonly evaluated through a sequence of invocations to highly optimised kernels provided in libraries such as BLAS and LAPACK. A sequence of kernels represents an algorithm, and in general, because of associativity, algebraic identities, and multiple kernels, one expression can be evaluated via many different algorithms. These algorithms are all mathematically equivalent (i.e., in exact arithmetic, they all compute the same result), but often differ noticeably in terms of execution time. 
When faced with a decision, high-level languages, libraries, and tools such as Julia, Armadillo, and Linnea choose by selecting the algorithm that minimises the FLOP count.
In this paper, we test the validity of the FLOP count as a discriminant for dense linear algebra algorithms, analysing "anomalies": problem instances for which the fastest algorithm does not perform the least number of FLOPs.

To do so, we focused on relatively simple expressions and analysed when and why anomalies occurred.
We found that anomalies exist and tend to cluster into large contiguous regions.
For one expression anomalies were rare, whereas for the other they were abundant.
We conclude that FLOPs is not a sufficiently dependable discriminant even when building algorithms with highly optimised kernels.
Plus, most of the anomalies remained as such even after filtering out the inter-kernel cache effects.
We conjecture that combining FLOP counts with kernel performance models will significantly improve our ability to choose optimal algorithms.

\end{abstract}

\begin{CCSXML}
<ccs2012>
   <concept>
       <concept_id>10002950.10003705</concept_id>
       <concept_desc>Mathematics of computing~Mathematical software</concept_desc>
       <concept_significance>500</concept_significance>
       </concept>
   <concept>
       <concept_id>10002950.10003705.10011686</concept_id>
       <concept_desc>Mathematics of computing~Mathematical software performance</concept_desc>
       <concept_significance>500</concept_significance>
       </concept>
   <concept>
       <concept_id>10010147.10010148.10010149.10010158</concept_id>
       <concept_desc>Computing methodologies~Linear algebra algorithms</concept_desc>
       <concept_significance>500</concept_significance>
       </concept>
 </ccs2012>
\end{CCSXML}

\ccsdesc[500]{Mathematics of computing~Mathematical software}
\ccsdesc[500]{Mathematics of computing~Mathematical software performance}
\ccsdesc[500]{Computing methodologies~Linear algebra algorithms}

\keywords{linear algebra, algorithm selection, scientific computing}


\maketitle

\section{Introduction}

Operations that manipulate matrices and vectors are known as linear algebra expressions.
These expressions,
such as $(X^T X)^{-1} X^T y$, are the cornerstone of countless applications and programs in both science and industry. 
Examples of convoluted linear algebra expressions can be found in, e.g., 
image restoration~\cite{tirer2018image}, 
information theory~\cite{albataineh2014blind,hejazi2015robust}, and 
signal processing~\cite{ding2016sparsity,rao2017robust}.
Oftentimes, the time to evaluate these expressions can have a substantial impact on the overall performance of the application~\cite{gonzalez2008effect}.
For this reason, the problem of how to translate linear algebra expressions into efficient code is central in HPC~\cite{abdelfattah2021set}.
The code to evaluate an expression is not unique.
In fact, there might be a myriad of mathematically equivalent algorithms.
This paper is a step towards improving the selection of the fastest algorithm for a given expression.

The manual translation of an expression into code is a task that requires thorough knowledge of both numerical linear algebra and the target computer architecture.  
The complexity quickly multiplies when the code is to be portable to different architectures or when the sizes of the operands are unknown at the time of translation.
Over the past several decades, the numerical linear algebra community has invested remarkable effort into identifying and optimising a relatively small set of basic kernels that can be used as building blocks when crafting code to evaluate more complex linear algebra expressions. 
The BLAS (Basic Linear Algebra Subprograms)~\cite{dongarra1985proposal,dongarra1990set,lawson1979basic} and LAPACK (Linear Algebra PACKage)~\cite{anderson1999lapack} libraries are two prominent results of this effort. 
The numerous alternative implementations of the BLAS and LAPACK collectively provide high performance on a multitude of different computer architectures, ranging from single-core to shared-memory multicore systems, and to accelerators such as GPUs.

The availability of such portable high-performance libraries shifts the problem of translating an expression into code away from optimising architecture-specific code into mapping the expression to a sequence of kernel calls.
These sequences of kernel calls, which might include bits between calls to transform data structures, is what we will henceforth refer to as \emph{algorithms}.
The problem of translating an expression to an algorithm while minimising a cost function is known as the Linear Algebra Mapping Problem (LAMP)~\cite{psarras2019linear}.
However, even the task of enumerating the set of mathematically equivalent algorithms for a given expression can be quite complex. 
As an example, consider the following expression used in signal processing~\cite{ding2016sparsity}:
\begin{equation*}\label{eqn::genLeast}
% \[
x := \left(A^{-T} B^T B A^{-1} + R^T LR\right)^{-1} A^{-T}B^T B A^{-1} y.
% \]
\end{equation*}
Tools such as Linnea~\cite{barthels2021} reveal that there are hundreds of algorithms that evaluate this expression by a sequence of calls to kernels in BLAS and LAPACK. 
These algorithms, while mathematically equivalent, can and often do have different FLOP counts and execution times. 
The problem, then, boils down to selecting an algorithm that is reliably among the fastest ones from this set for a specific combination of library implementations and computer architecture.
\fix{Many widely-used programming languages face this problem and advances in algorithm selection will benefit their users.}

Since the execution time is not known until after the computation has been performed, and accurate performance prediction remains a challenging problem~\cite{iakymchuk2012modeling,peise2012performance} (even more so for sequences of kernel calls~\cite{peise2014study}) we are forced to make educated guesses based on proxies related to the execution time.
A simple and natural strategy is to select one of the cheapest algorithms, i.e., an algorithm with the minimum floating point operation (FLOP) count.
In fact, this strategy is currently used by the tool Linnea~\cite{barthels2021}, and the library Armadillo~\cite{sanderson2016armadillo} as well as compilers for the programming language Julia are also known to employ this FLOP count strategy for certain types of linear algebra expressions.

Selecting algorithms based on FLOP count is likely good enough to avoid very costly algorithms.
For example, if $A$ is an $n \times n$ matrix and $x, y$ are vectors of length $n$, then the algorithm implied by the parenthesisation
$(x y^T) A$
involves a vector outer product costing $n^2$ FLOPs and a matrix multiplication costing $2n^3$ FLOPs.
The algorithm implied by $x (y^T A)$, on the other hand, involves two matrix-vector multiplications costing $2n^2$ FLOPs each.
However, when the difference in FLOP count between algorithms is closer to zero, the performance (FLOP/s) of the various kernels as well as caching effects between kernel calls progressively gain importance. 
We refer to instances for which one of the cheapest algorithms is not one of the fastest algorithms as \emph{anomalies}.
While it is well-known that anomalies exist, less is known about how abundant anomalies are and which key factors explain their presence. 

In this paper, we experimentally study the abundance and distribution of anomalies in expressions of the form $A B C D$ and $A A^T B$ and seek to explain their presence.
Being able to identify regions of the problem space where one can and cannot rely on FLOPs as a discriminant is especially important when some matrix sizes are unknown. 
This work is an initial investigation in this direction.
The expression $A B C D$, known as "matrix chain", originates six distinct algorithms that all rely on only one BLAS kernel (GEMM); this kernel is unquestionably the most studied and optimised numerical kernel in any area of computing. 
We conjecture that if anomalies can be found in this expression, then they will be even more abundant in more complex expressions whose algorithms use multiple kernels. 
The expression $A A^T B$ was chosen since it is one of the simplest operations with algorithms constructed from three different BLAS kernels (GEMM, SYRK, SYMM). 
This paper makes the following contributions.
\begin{itemize}
    \item \fix{We show that anomalies occur even when selecting algorithms built from optimised libraries for relatively simple expressions regardless of the combination of library implementation and computing platform.}
    \item We experimentally investigate both the abundance and severity of anomalies. In doing so, we present empirical evidence that anomalies can be both abundant ($\approx10\%$) and severe ($45\%$ more FLOPs but $40\%$ lower execution time) even for simple expressions such as $A A^T B$.
    \item We present empirical evidence that the interplay between the shapes of the performance profiles of the kernels is a major explanation of anomalies.
    \item We present empirical evidence that FLOP counts combined with performance profiles of kernels may be able to predict a large fraction of anomalies.
    \item With this study, we make a necessary step towards solving the LAMP with symbolic sizes.
\end{itemize}

The remainder of the paper is organised as follows. 
Related work is described in Section~\ref{sec:related}.
In Section~\ref{sec:methodology}, we describe the experimental methodology.
The results of the experiments are presented in Section~\ref{sec:results}.
Finally, conclusions and future work are described in Section~\ref{sec:conclusion}.

\section{Related Work}\label{sec:related}

%%%%%%%%%%%%%%%%%%%%%
% Introduction part of related work. 
%%%%%%%%%%%%%%%%%%%%%
Before the introduction of CPU caches, the cost of a numerical algorithm was effectively proportional to the FLOP count --- optimising execution time meant minimising FLOP count. 
This is a major reason why significant effort has been put into developing algorithms that minimise the FLOP count (e.g., \cite{vetterli1984simple,strassen1969gaussian,dhillon1997new,xu2008fast}).
By introducing CPU caches, the paradigm to reduce execution time radically changed. 
The execution time for an operation not only depends on the FLOP count but also on the location of the operands in the memory hierarchy and how these are accessed~\cite{dongarra1984implementing,dongarra1986squeezing}.
When algorithms are tailored to modern computer architectures with deep memory hierarchies, the importance of also optimising for caching becomes apparent.
Indeed, there are many examples where an \emph{increase} in FLOP count results in a \emph{decrease} in execution time (e.g., \cite{bischof1987wy,bischof1994parallel,lang1998using,buttari2008parallel}).

We stress that the FLOP count may in principle be a good discriminant between algorithms without being an accurate predictor of execution time. 
Indeed, being able to select a fastest algorithm from a finite set of algorithms does not imply an ability to accurately predict any one algorithm's execution time~\cite{sankaran2020discriminating}.

Algorithm selection followed by automatic code generation has been heavily applied in the area of linear signal transforms (such as FFT), which are very specific linear algebra expressions.
FFTW~\cite{frigo1998fftw} is a widely used library for fast transforms that uses this approach.
The library generates many different \emph{codelets} that it combines into many different \emph{plans} (algorithms).
For a given instance, the library estimates the execution time of plans through empirical performance testing.
SPIRAL~\cite{franchetti2018spiral} is a program generation software for linear transforms and other mathematical functions. 
For each transform there is a set of rules for manipulating the input, resulting in a set of algorithms. 
The algorithm selection is determined by a cost function, which can be FLOP count, accuracy, instruction count, or execution time (default). 
The execution time is determined by empirical performance testing.

%%%%%%%%%%%%%%%%%%%%%%%
% Part focusing on the LAMP. 
%%%%%%%%%%%%%%%%%%%%%%
Algorithm selection and automatic code generation has also, but to a lesser extent, been used to solve the LAMP~\cite{psarras2019linear}.
The Transfor~\cite{gomez1998maple} Maple package is one of the earliest attempts to translate linear algebra expressions into BLAS kernel calls.
The translation process generates a single algorithm for a given input expression based on a set of handcrafted rules.
Hence, there is no proper algorithm selection phase. 
The selection is implicit in the design of the rules and the code generator. 
The rules were primarily designed to limit the amount of auxiliary memory.
The work in~\cite{mcfarlin2007library} presents a source-to-source compiler that converts linear algebra expressions in Octave into calls to library functions (e.g., BLAS kernels).
The code is first flattened into a sequence of primitive Octave operations.
Then individual statements are, whenever possible, translated into an equivalent library function call. 
In cases where the mapping from operation to library function is not unique, the best choice for the target platform is made on the basis of empirical performance testing done ahead of time.
% Whenever possible, sequences of primitive operations are subsumed into one library function call. 
Linnea~\cite{barthels2021} is a more recent attempt to solve the LAMP.
This tool targets both BLAS and LAPACK and outputs Julia code.
Linnea generates algorithms by rewriting the input expression and identifying parts of it that are computable by BLAS/LAPACK kernels. 
The algorithm selection is done by minimising the FLOP count with a best-first search.
High-level languages and libraries (e.g., \cite{sanderson2016armadillo,Matlab2022,eigenweb,bezanson2018julia,harris2020array,gnuoctave,Rlang}) also face the LAMP. 
Some have been shown to generate solutions to the LAMP that are worse than the solution obtained by minimising the FLOP count~\cite{psarras2019linear}.

\section{Methodology}\label{sec:methodology}

We aim to demonstrate the existence of anomalies, estimate their abundance, quantify their severity, and identify major factors that explain their presence. 
The experimental study of anomalies is complicated, since their presence clearly depends on the specific expression, the set of algorithms considered for the expression, and the computer system (hardware, software, and configuration) --- each with infinitely many options.
Moreover, the set of instances for a given expression is also infinite.

The algorithms in this paper make use of three BLAS kernels, which we briefly introduce and give an estimated FLOP count of in Section~\ref{sec:kernels-flops}.
The expressions we experiment on and the set of algorithms we consider for each are introduced in Section~\ref{sec:expressions}. 
The procedure for classifying instances as anomalies is described in Section~\ref{sec:anomaly-classification} along with two scores we use to measure their severity.
The design of the experiments is described in Section~\ref{sec:experiments}.

\subsection{Kernels and their FLOP Counts}\label{sec:kernels-flops}

The algorithms are constructed almost exclusively from calls to these three kernels in the BLAS library:
\begin{itemize}
    \item The GEMM kernel (as used here) computes matrix products of the form $A B$. 
    If the size of $A$ is $m \times k$ and $B$ is $k \times n$, then we take the FLOP count to be $2 m n k$.

    \item The SYRK kernel (as used here) computes one triangle of symmetric matrix products of the form $A A^T$. 
    If the size of $A$ is $m \times k$, then we take the FLOP count to be $(m+1) m k$.

    \item The SYMM kernel (as used here) computes matrix products of the form $A B$ where $A$ is symmetric. 
    If the size of $A$ is $m \times m$ and $B$ is $m \times n$, then we take the FLOP count to be $2 m^2 n$.
\end{itemize}
\fix{We do not consider algorithms that replace a call to a level~3 BLAS kernel with an equivalent sequence of calls to lower-level BLAS kernels, since performance will degrade if the BLAS library is reasonably optimised.}

\begin{figure}[tbh]
    \centering
    \includegraphics[width=7cm]{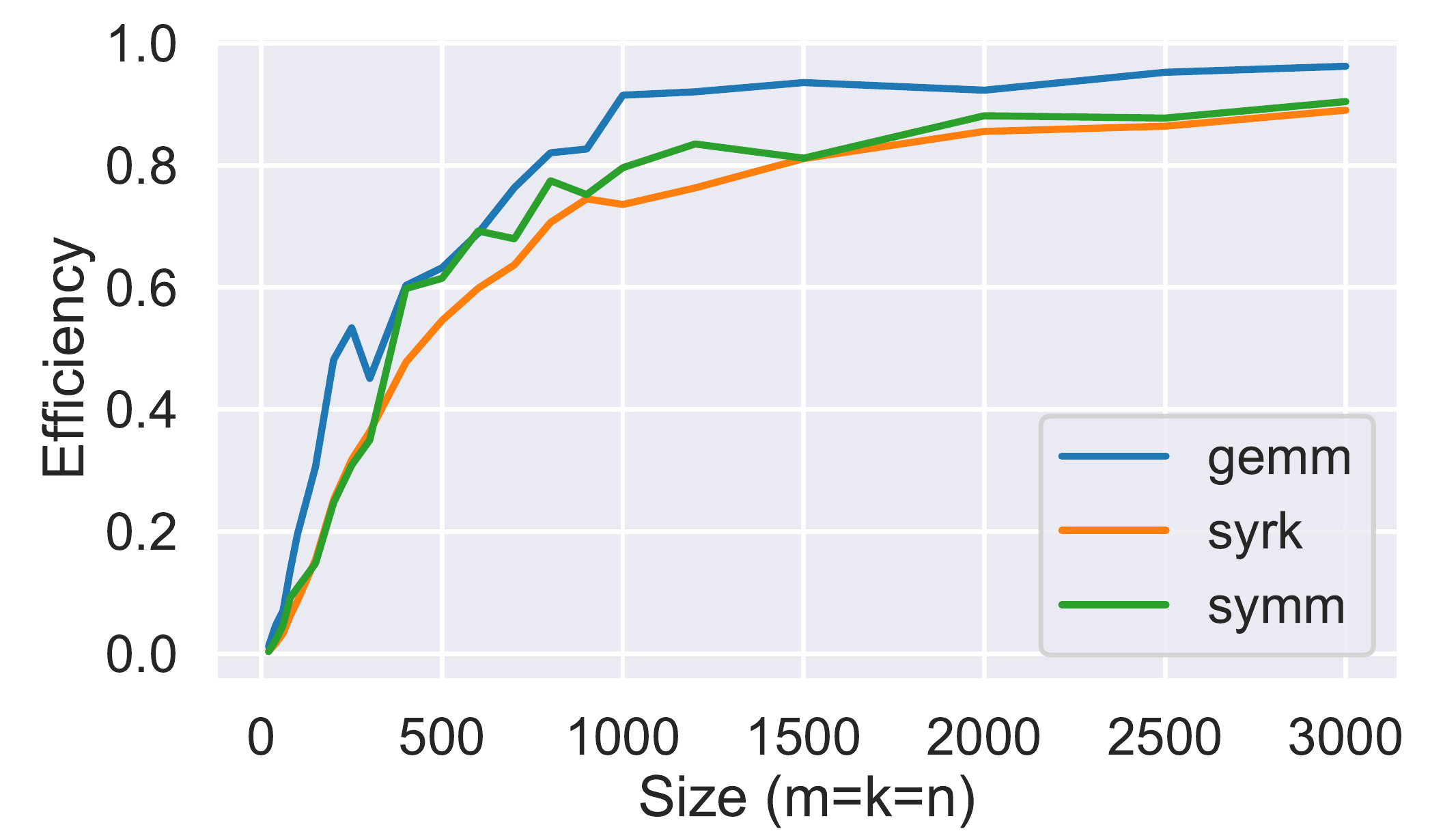}
    \caption{Efficiency of the BLAS kernels GEMM, SYRK, and SYMM as the size of the operands (all square matrices) grows. }
    \label{fig::eff_kernel}
\end{figure}

\fix{\emph{Efficiency} (of a kernel or algorithm) is defined as the ratio of the measured performance to the computer's theoretical peak performance.}
In Figure~\ref{fig::eff_kernel}, we show the efficiencies of the kernels as a function of the size of the input operands for the case where all operands are square matrices. The differences are small but noticeable.

\subsection{Expressions and their Algorithms}\label{sec:expressions}

\subsubsection{The Matrix Chain Expression}\label{sec:matrix-chain}

\begin{figure}[tbh]
    \centering
    \includegraphics[width=\columnwidth]{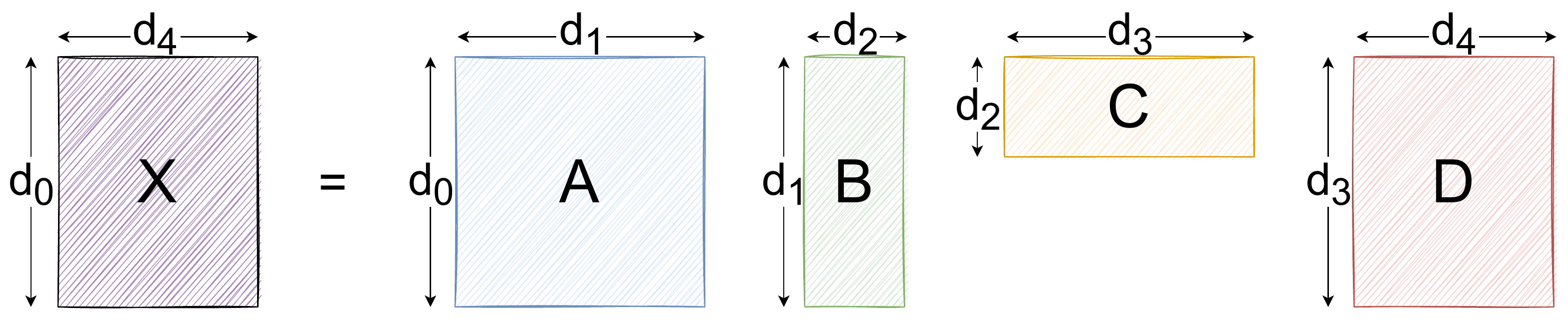}
    \caption{Illustration of the matrix chain $A B C D$.}
    \label{fig::MC4_layout}
\end{figure}

The \emph{matrix chain expression} we consider takes the form
\begin{equation*}\label{eqn::matrixchain}
X := A B C D,
\end{equation*}
where $A\in\mathbb{R}^{d_0\times d_1}$, $B\in\mathbb{R}^{d_1\times d_2}$, $C\in\mathbb{R}^{d_2 \times d_3}$, and $D \in\mathbb{R}^{d_3\times d_4}$ (see Figure~\ref{fig::MC4_layout}).
The matrices are assumed to be dense and unstructured.
Hence, only their sizes (not their elements) affect performance. 
An instance of this expression is specified by the tuple $(d_0,d_1,d_2, d_3,d_4)$. 

\begin{figure}[htbp]
    \centering
    \includegraphics[width=\columnwidth]{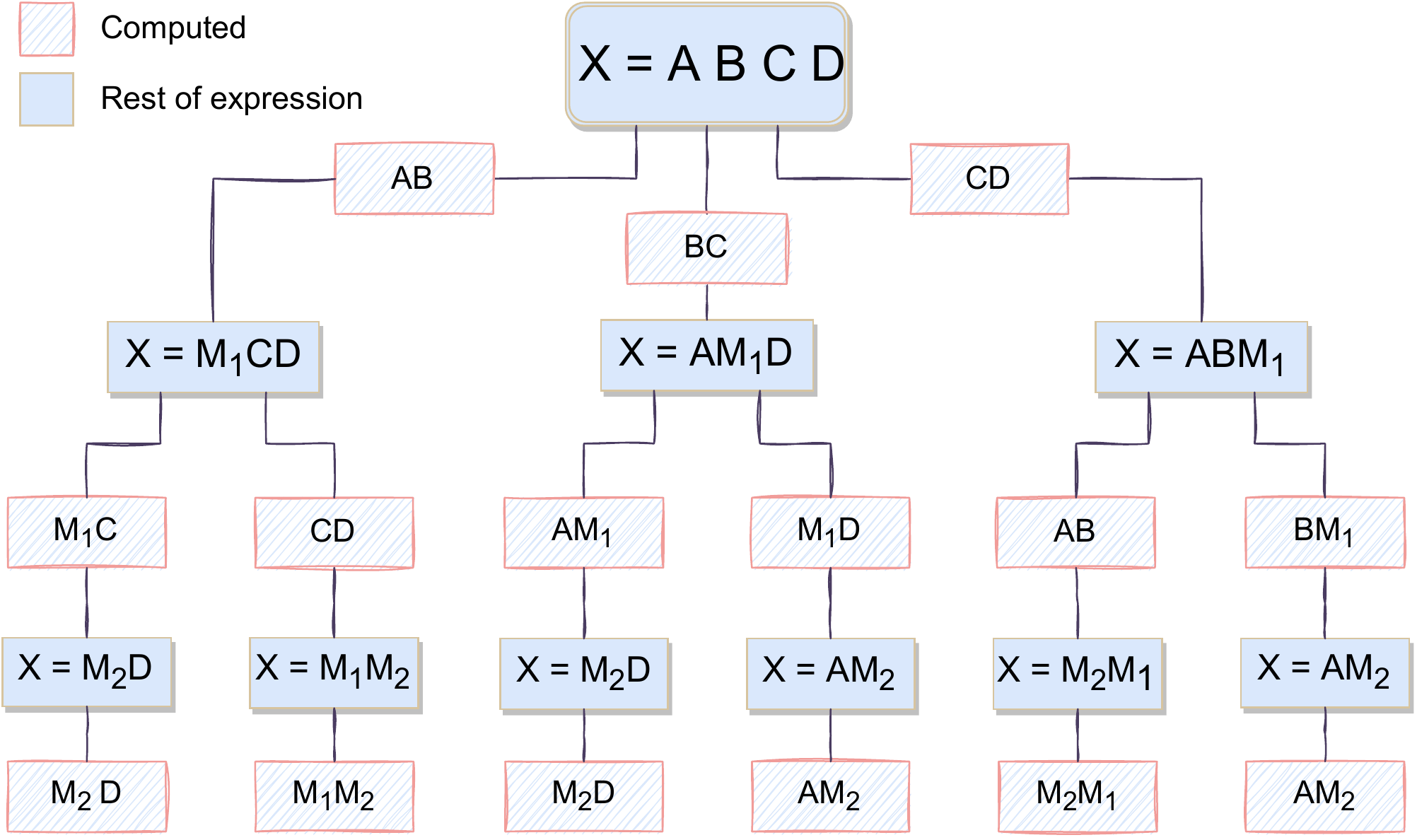}
    \caption{
        The algorithms for the matrix chain expression. 
        Each path from the root node (top) to a leaf node represents an algorithm.
    }
    \label{fig::MC4_algs}
\end{figure}

The set of algorithms for the matrix chain expression is taken to be all (reasonable) sequences of calls to the BLAS kernel GEMM that evaluate the expression. 
The expression has three matrix multiplications and each GEMM performs one of them.
Due to associativity, the multiplications can be done in any order.
It follows that there are $3! = 6$ different but mathematically equivalent algorithms, all illustrated in Figure~\ref{fig::MC4_algs}.
Specifically, the algorithms and their FLOP counts are as follows.
\begin{itemize}
    \item Algorithm~1: $M_1 := A B$; $M_2 := M_1 C$; $M_2 D$ with a FLOP count of 
    $2 d_0 ( d_1 d_2 + d_2 d_3 + d_3 d_4 )$. 
    \item Algorithm~2: $M_1 := A B$; $M_2 := C D$; $M_1 M_2$ with a FLOP count of 
    $2 d_2 ( d_0 d_1 + d_0 d_4 + d_3 d_4 )$. 
    \item Algorithm~3: $M_1 := B C$; $M_2 := A M_1$; $M_2 D$ with a FLOP count of 
    $2 d_3 ( d_0 d_1 + d_0 d_4 + d_1 d_2 )$.
    \item Algorithm~4: $M_1 := B C$; $M_2 := M_1 D$; $A M_2$ with a FLOP count of 
    $2 d_1 ( d_0 d_4 + d_2 d_3 + d_3 d_4 )$.
    \item Algorithm~5: $M_1 := C D$; $M_2 := A B$; $M_2 M_1$ with a FLOP count of 
    $2 d_2 ( d_0 d_1 + d_0 d_4 + d_3 d_4 )$, same as Algorithm~2. 
    \item Algorithm~6: $M_1 := C D$; $M_2 := B M_1$; $A M_2$ with a FLOP count of 
    $2 d_4 ( d_0 d_1 + d_1 d_2 + d_2 d_3 )$. 
\end{itemize}

\subsubsection{The Expression $A A^T B$}\label{sec:aatb}

\begin{figure}[tbh]
    \centering
    \includegraphics[width=\columnwidth]{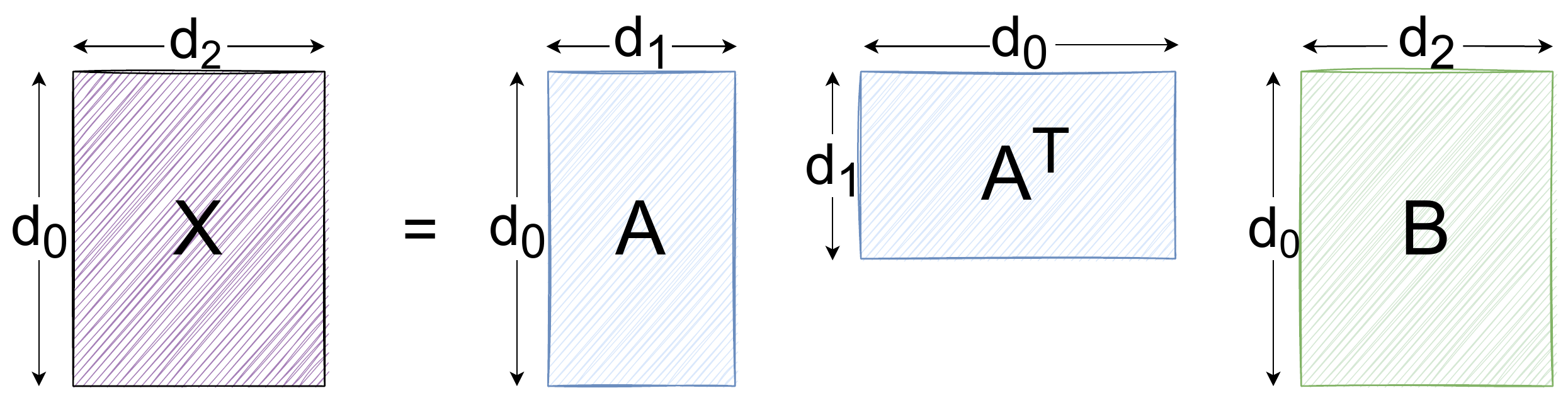}
    \caption{Illustration of the expression $A A^T B$.}
    \label{fig::aatb_layout}
\end{figure}

The second expression takes the form
\begin{equation*}\label{eqn::aatb}
X := A A^{T} B,
\end{equation*}
where $A \in\mathbb{R}^{d_0 \times d_1}$ and $B \in\mathbb{R}^{d_0 \times d_2}$. 
Since the matrices are assumed dense and unstructured, only the sizes of $A$ and $B$ (not their elements) affect performance. 
An instance of this expression is specified by the tuple $(d_0,d_1,d_2)$. 

We take the set of algorithms to be all (reasonable) combinations of the GEMM, SYRK, and SYMM kernels in the BLAS. 
Each kernel performs one of the two matrix multiplications in the expression, so each algorithm has two kernel calls.
There are $2! = 2$ ways to order the multiplications, but, as we will see, the choice of kernels gives rise to five different algorithms.

Suppose that $M := A A^T$ is done first. 
There are two kernels to choose from for each multiplication, resulting in four different algorithms. 
Since $M$ is symmetric, either GEMM or SYRK can be used. 
If SYRK is used, then only one triangle (the upper or lower) of $M$ is available.
Regardless, the subsequent multiplication $M B$ can be done using either GEMM or SYMM. 
If GEMM is used, then the triangle computed by SYRK must first be extended to a full matrix by copying its elements to fill in the missing triangle. 

Now suppose that $M := A^T B$ is done first. 
Only GEMM can be used. 
The subsequent multiplication with $A$ must also use GEMM.
Hence, there is only one algorithm in this case.

\begin{figure}[tbh]
    \centering
    \includegraphics[width=6cm]{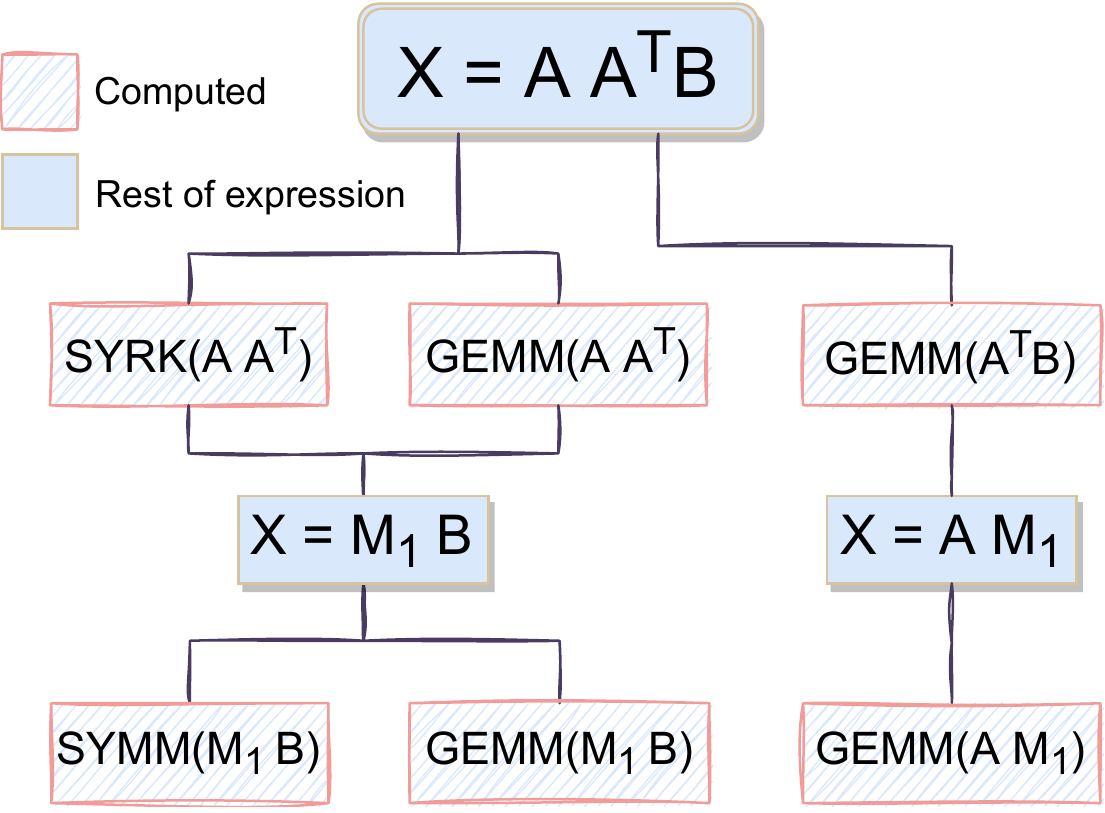}
    \caption{
        The algorithms for the expression $A A^T B$. 
        Each path from the root node (top) to a leaf node represents an algorithm.
        }
    \label{fig::aatb_algorithms}
\end{figure}

All in all, there are five algorithms, all illustrated in Figure~\ref{fig::aatb_algorithms}.
The algorithms and their FLOP counts are as follows.
\begin{itemize}
    \item Algorithm~1: SYRK for $M := A A^T$ followed by SYMM for $M B$. The FLOP count is taken to be $d_0 ((d_0 + 1)d_1 + 2 d_0 d_2)$.
    \item Algorithm~2: SYRK for $M := A A^T$ followed by GEMM for $M B$. The triangle computed by SYRK is copied to form a full matrix before the GEMM. The FLOP count is taken to be $d_0 ((d_0 + 1)d_1 + 2 d_0 d_2)$, the same as for Algorithm~1.
    \item Algorithm~3: GEMM for $M := A A^T$ followed by SYMM for $M B$. The FLOP count is taken to be $2 d_0^2 (d_1 + d_2)$.
    \item Algorithm~4: GEMM for $M := A A^T$ followed by GEMM for $M B$. The FLOP count is taken to be $2 d_0^2 (d_1 + d_2)$, the same as for Algorithm~3. 
    \item Algorithm~5: GEMM for $M := A^T B$ followed by GEMM for $A M$. The FLOP count is taken to be $4 d_0 d_1 d_2$.
\end{itemize}

\subsection{Anomaly Classification}\label{sec:anomaly-classification}

%Here we describe the procedure for anomaly classification of instances of expressions.  
The algorithms with the shortest execution time for the instance are labelled as the \emph{fastest} algorithms. 
The algorithms with the lowest FLOP count for the instance are labelled as the \emph{cheapest} algorithms.
An instance of an expression is classified as an anomaly if these two sets are disjoint, i.e., none of the cheapest algorithms are among the fastest. 

% Definition of time scores and FLOP scores.
To quantify the severity of an anomaly, we calculate two scores for each anomaly: {time score} and {FLOP score}. 
The \emph{time score} is defined as 
\begin{equation*}\label{eqn::time_score}
\frac{T_{\rm cheapest} - T_{\rm fastest}}{T_{\rm cheapest}} \in [0\%, 100\%],
\end{equation*}
where $T_{\rm cheapest}$ is the shortest execution time amongst the set of cheapest algorithms and $T_{\rm fastest}$ is the shortest execution time amongst all algorithms. 
A time score of $x\%$ means that the fastest algorithm has an $x\%$ shorter execution time than that of the fastest amongst the cheapest algorithms.
If an instance is an anomaly, then $x > 0\%$ and otherwise, $x = 0\%$.
The larger the time score, the more severe the anomaly is considered to be. 

The \emph{FLOP score} is defined as
\begin{equation*}\label{eqn::flops_score}
\frac{F_{\rm fastest} - F_{\rm cheapest}}{F_{\rm fastest}} \in [0\%, 100\%],
\end{equation*}
where $F_{\rm cheapest}$ is the FLOP count of the cheapest algorithms and $F_{\rm fastest}$ is the FLOP count of the cheapest amongst the fastest algorithms. 
A FLOP score of $x\%$ means that the cheapest algorithms perform $x\%$ fewer FLOPs than the fastest algorithm. 

% Thresholds.
In practice, the fastest of the cheapest algorithms could potentially be just barely slower than the fastest algorithm overall and, thus, the instance would be classified as an anomaly even though this would hardly be a significant distinction in practice. 
Therefore, in all experiments we require a time score above a certain threshold (usually $10\%$; see Section~\ref{sec:results}) before classifying an instance as an anomaly.

% Experiment to find anomalies.
\subsection{Experiments}\label{sec:experiments}

Intuitively, anomalies should be expected to cluster together into contiguous regions since execution times of nearby instances tend to be similar due to the following observations.
First, the FLOP counts of the kernels are continuous functions of the sizes of the matrices (see Section~\ref{sec:kernels-flops}).
Second, a small change in size in any operand of a kernel call typically induces a small change in the performance of said kernel (see Figure~\ref{fig::eff_kernel}).
The execution time for the instance $(d_0, d_1, d_2)$ is thus expected (but not guaranteed) to be similar to the execution time for nearby instances.

We carried out three experiments on both expressions using double precision (64 bit) arithmetic. 
The first experiment (Section~\ref{sec:exp-1}) is a random search for anomalies to estimate their abundance and severity. 
The second experiment (Section~\ref{sec:exp-2}) explores the vicinity of the anomalies from the first experiment to determine to what extent anomalies cluster into contiguous regions. 
The third experiment (Section~\ref{sec:exp-3}) aims to determine how many of the anomalies found in the second experiment could have been predicted from performance profiles of the three kernels obtained via separated benchmarks. 

% Choice of computer system.
All experiments were conducted on a Linux-based system with 40~GB of RAM and an Intel Xeon Silver 4210 processor, which contains ten physical cores. 
Turbo Boost was enabled. 
All ten physical cores were used, and threads were pinned to ensure one thread per physical core. 
The source code, available on GitHub\footnote{\href{https://github.com/FranLS7/LAMB}{https://github.com/FranLS7/LAMB}}, was compiled with GCC\footnote{GCC version 7.5.0} and linked against the Intel Math Kernel Library\footnote{MKL version 2019.0.5}. 

% Noise handling.
To reduce the impact of measurement noise, each test (of a specific algorithm on a specific instance) was repeated ten times and the median was recorded as the execution time. 
To eliminate cache effects, the cache was flushed prior to each repetition.

\subsubsection{Experiment 1: Random search}\label{sec:exp-1}

The first experiment searches for anomalies using random search.
Instances are repeatedly sampled uniformly at random with replacement from a specified search space (see Section~\ref{sec:results}).  
All algorithms are tested on the instance and the instance is then classified as an anomaly or not (see Section~\ref{sec:anomaly-classification}). 
If the instance is an anomaly, then its time score and FLOP score are recorded.

% Experiment to explore regions.
\subsubsection{Experiment 2: Lines through regions}\label{sec:exp-2}

Mapping entire regions proved to be computationally infeasible.
Instead, in this experiment, the regions around all the anomalies found in \hyperref[sec:exp-1]{Experiment~1} are intersected with axis-aligned lines in all dimensions through the original anomaly.
For example, if $(d_0, d_1, d_2)$ is an anomaly from \hyperref[sec:exp-1]{Experiment~1},
then the three axis-aligned lines are: $(d_0 \pm 10 x, d_1, d_2)$ and $(d_0, d_1 \pm 10 x, d_2)$ and $(d_0, d_1, d_2 \pm 10 x)$ for $x = 0, 1, 2, \ldots$.
The line is traversed (in steps of 10 to make the experiment feasible) in both directions 
from the original anomaly and stops either when the boundary of the search space is reached or some distance passed the end of the region. 

Transient changes in performance, system jitter, and measurement noise might cause an isolated instance inside a region to be classified as not an anomaly.
One or two consecutive instances classified as not anomalies are considered to be a hole in the region rather than marking the end of it. 
Thus, the end of a region is characterised by three or more consecutive instances classified as not anomalies.
The first of those three is labelled as the boundary of the region.
If the traversal reaches the boundary of the search space, then the last instance is labelled as the boundary of region.
Let $a$ and $b$ (with $a < b$) be the location of the boundary points of the region along the line.
Then $b - a - 1$ is referred to as the \emph{thickness} of the region in the dimension of the line.

% Experiment to filter out inter-kernel caching effects.
\subsubsection{Experiment 3: Prediction from benchmarks}\label{sec:exp-3}

The performance of an algorithm consisting of a sequence of kernel calls is determined by the FLOP counts and the observed performance of the individual calls. 
Since we flush the cache before each execution, the performance of the first call is expected to be close to the ideal (benchmarked) performance of the kernel. 
However, the state of the cache when exiting one call may affect the performance of the next and subsequent calls.

The third experiment determines to what extent anomalies could have been predicted from
benchmarked performance profiles of the kernels.
For each sample in \hyperref[sec:exp-2]{Experiment~2}, the algorithms collectively generate a small set of specific calls.
In this third experiment, all of these calls are benchmarked in isolation with a flushed cache. 
Thus, for each call made in \hyperref[sec:exp-2]{Experiment~2}, we obtain a benchmarked execution time. 
By summing over the calls in any given algorithm, we obtain a prediction of the execution time of the algorithm based on benchmarked performance of the kernels.

\begin{figure}[tbh]
    \centering
    \includegraphics[width=8.3cm]{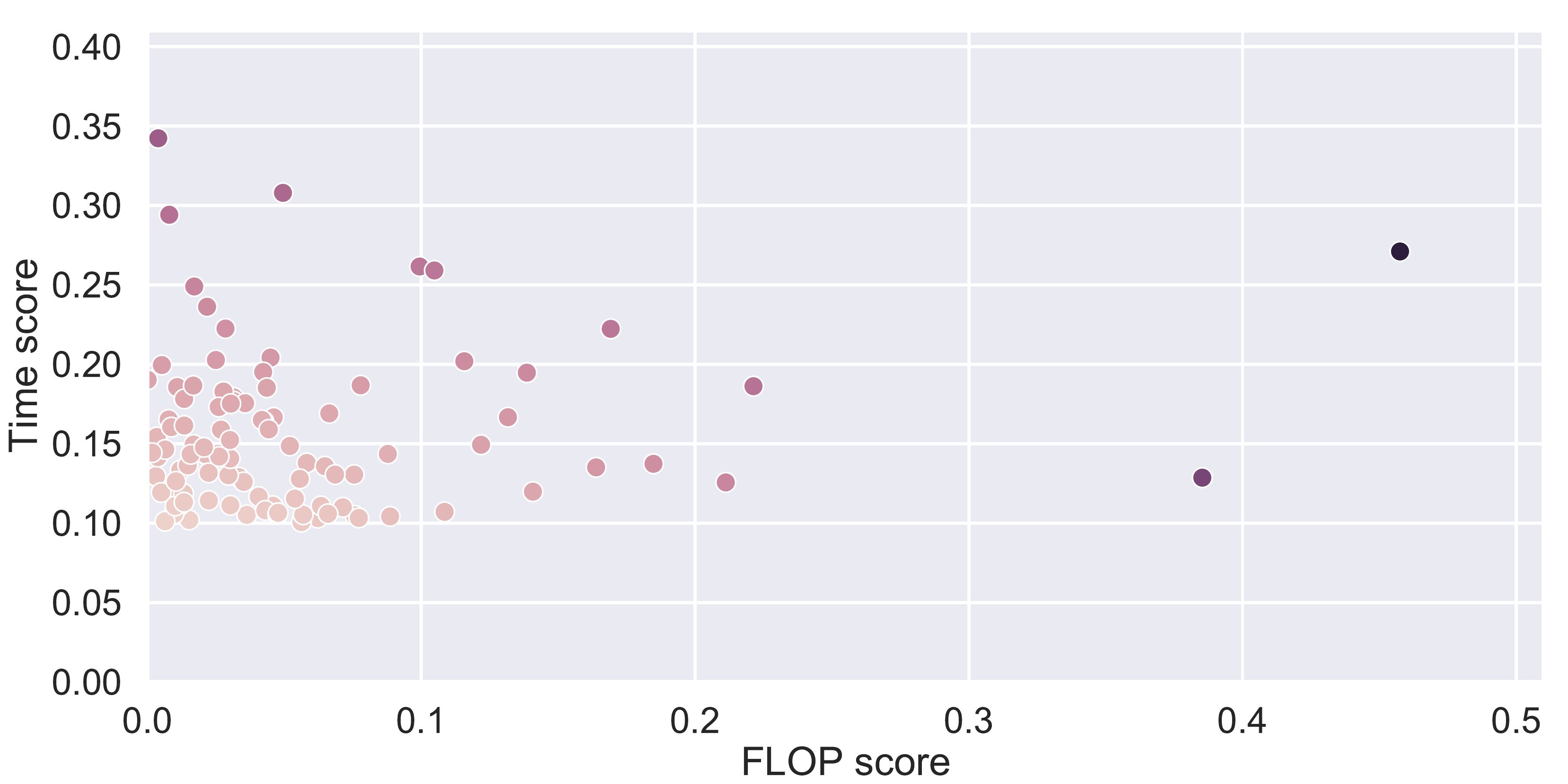}
    \caption{Scatter plot of the time score versus FLOP score for 100 anomalies of the matrix chain expression found in \hyperref[sec:exp-1]{Experiment~1}.}
    \label{fig::MC4_timevsflops}
\end{figure}

\begin{figure*}[tbh]
    \centering
    \includegraphics[width=\textwidth]{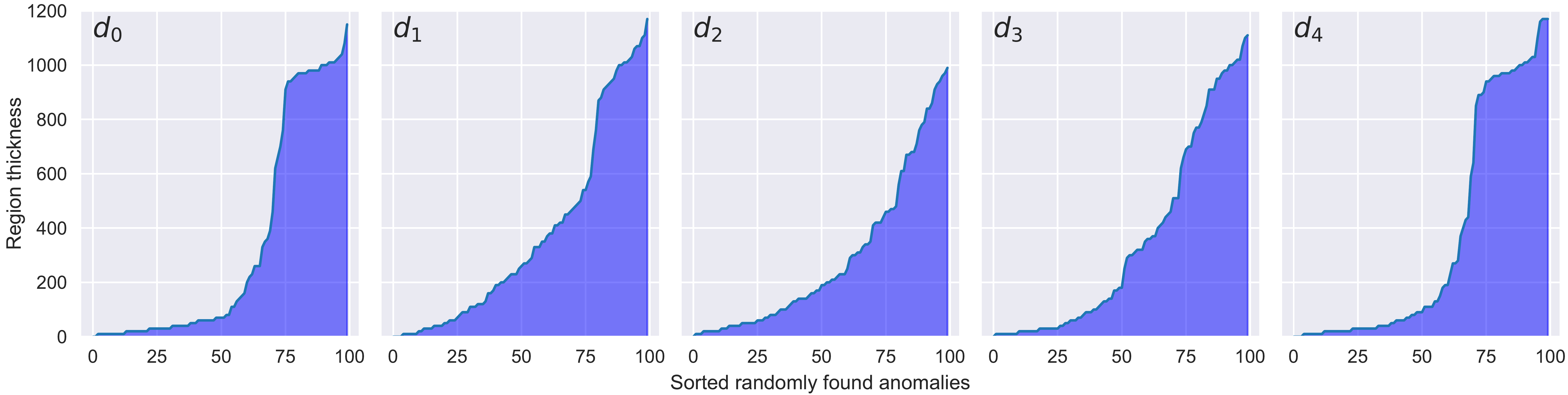}
    \caption{The distribution of the thicknesses of the regions around the 100 anomalies found in \hyperref[sec:exp-1]{Experiment~1} on the matrix chain expression \fix{in each dimension from $d_0$ (left) to $d_4$ (right).}}
    \label{fig::MC4_Dim_Expl}
\end{figure*}

\section{Results}\label{sec:results}
Section~\ref{sec:result-chain} presents results for the matrix chain expression $A B C D$, 
and Section~\ref{sec:result-other} presents corresponding results for the expression $A A^T B$.

\subsection{The Matrix Chain Expression}\label{sec:result-chain}

\subsubsection{Abundance}\label{sec:MC4_rand_anom}

The search space for \hyperref[sec:exp-1]{Experiment~1} was bounded by $20 \leq d_i \leq 1200$ for $i = 0, 1, \ldots, 4$.
The upper limit was chosen so that anomalies were plausible since GEMM's performance usually plateaus when the dimensions sizes exceed this limit (and therefore no anomalies could be found there). 
The search continued until 100 distinct anomalies were found. 
The time score threshold (see Section~\ref{sec:anomaly-classification}) was set to $10\%$.

Figure~\ref{fig::MC4_timevsflops} shows the result of the experiment. 
Finding 100 anomalies required 22,962 samples, which translates to an estimated abundance of $0.4\%$ in the constrained search space. 
Conversely, a cheapest algorithm is either the fastest or within $10\%$ of the fastest in $99.6\%$ of instances.
Most anomalies have a FLOP score below $10\%$ and a time score below $20\%$ and hence most of the (already rare) anomalies are not particularly severe. 
On the other hand, there are cases where an algorithm performing barely more than the minimum number of FLOPs is $35\%$ faster than the cheapest algorithms.

\subsubsection{Regions}
The time score threshold for \hyperref[sec:exp-2]{Experiment~2} was set at $5\%$. 
Figure~\ref{fig::MC4_Dim_Expl} shows the distribution of the thicknesses of the anomalous regions in each of the five dimensions $d_0$ (left) through $d_4$ (right). 
The maximum thickness is close to 1181 since that is the number of instances in a single line from 20 to 1200.

\subsubsection{Region Boundaries}\label{sec:mc4_dim_expl}

\begin{figure}[tbh]
    \centering
    \includegraphics[width=\columnwidth]{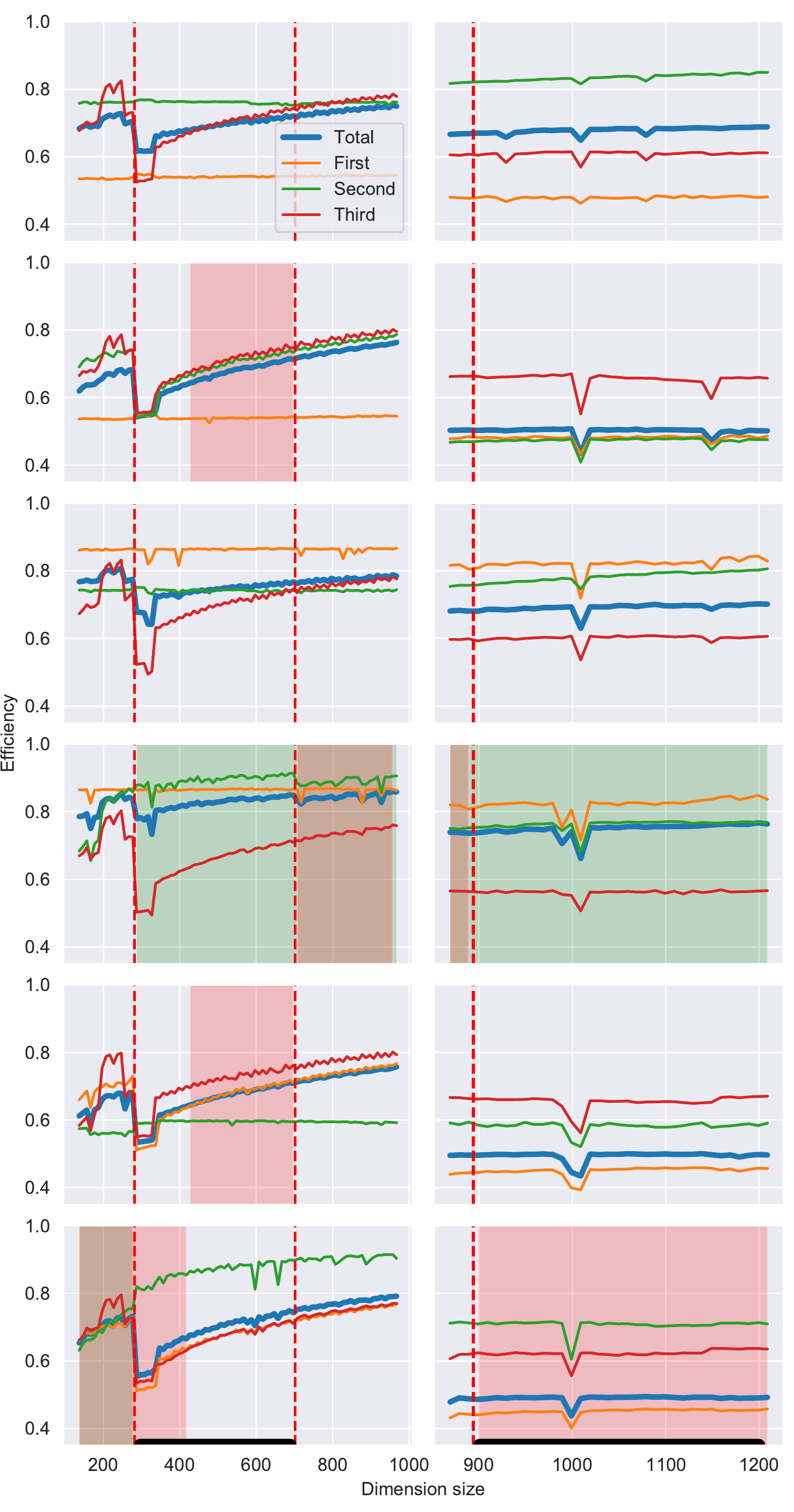}
    \caption{\fix{Two examples (left and right) of efficiencies in anomalous regions demonstrating the two types of transitions.
    Each row corresponds to one of the six algorithms;
    the columns correspond to a different anomaly (and dimension traversed).
    Left: Efficiency along the line $(331,279,338,854,427 \pm 10 x)$, dimension $d_4$ being traversed.
    Right: Efficiency along the line $(320,172,293,919 \pm 10 x,284)$, dimension $d_3$ being traversed.}
    }
    \label{fig::mc4_eff_dim}
\end{figure}

The data from \hyperref[sec:exp-2]{Experiment~2} allows for inspection of what happens near the boundaries of the regions. 
Figure~\ref{fig::mc4_eff_dim} shows how the efficiency of the six algorithms varies along one of the lines for two of the anomalies found in \hyperref[sec:exp-1]{Experiment~1}. 
When a line crosses a region boundary, either the efficiency of one or more algorithms abruptly changes or the efficiencies of all algorithms change non-abruptly.
\fix{Abrupt changes in kernel efficiency are likely caused by an internal change in the underlying chosen algorithmic variant.}
These two types of transitions are complementary and hence there is no third type.

The two examples in Figure~\ref{fig::mc4_eff_dim} serve to illustrate the two types of transitions.
Each of the six rows corresponds to an algorithm. 
In addition to the efficiency of the whole algorithm (solid blue), we plot the efficiency of the three calls to GEMM (other colours) involved in each algorithm. 
The background colour indicates for each sample along the line i) in red a cheapest algorithm, ii) in green a fastest algorithm, and iii) in brown a cheapest \emph{and} fastest algorithm.
\fix{Therefore, the segments with brown background colour are not anomalous, whereas those with green and red are anomalous.}
The region boundaries are marked with vertical red dashed lines.
The region's extension in the explored dimension is marked with a black line on the horizontal axis (coinciding with the anomalous areas -- where no algorithm presents a brown background).

Consider the anomaly in the left column. 
As $d_4$ increases along the line, Algorithm~6 is both the cheapest and the fastest up until approximately $300$. 
For larger values of $d_4$, Algorithm~6 is still the cheapest but Algorithm~4 is the fastest. 
When $d_4$ passes $410$, Algorithms~2 and~5 (same FLOP count) become the cheapest, but Algorithm~4 remains the fastest. 
When $d_4$ passes $700$, Algorithm~4 becomes both fastest and cheapest, marking the end of the region. 

Now consider the anomaly in the right column. 
For small values of $d_3$, Algorithm~4 is both cheapest and fastest, until $d_3$ reaches $900$ and the region starts. 
For larger values of $d_3$, Algorithm~4 continues to be the fastest but Algorithm~6 becomes the cheapest.

The transition at the boundary of an anomalous region is sometimes due to an abrupt change in efficiency in one or more kernels.
The transition near $300$ in the left column of Figure~\ref{fig::mc4_eff_dim} is an example of this type of transition.
If there is no abrupt change in efficiency, then the transition is more gradual. 
The transition near $700$ in the left column and the transition near $900$ in the right column are both examples of this type of transition.

\subsubsection{Prediction from Benchmarks}

The time score threshold for \hyperref[sec:exp-3]{Experiment~3} was set to $5\%$. 
For all instances sampled along a line, the anomaly classification derived from the measured execution time (\hyperref[sec:exp-2]{Experiment~2}) is taken as ground truth and the benchmark data (\hyperref[sec:exp-3]{Experiment~3}) is used to predict anomalies for the same set of instances. 
Thus, each instance falls into one of four categories (actual anomaly yes/no and predicted anomaly yes/no).
The result of \hyperref[sec:exp-3]{Experiment~3} is shown in the form of a confusion matrix in Table~\ref{tbl::CF_MC4}. 
Approximately $92\%$ of the anomalies could have been predicted from only the benchmark data, and $96\%$ of the anomalies predicted by the benchmark data were actual anomalies. 

\makegapedcells
\begin{table}[htbp]
\begin{tabular}{cc|cc|c}
\multicolumn{2}{c}{}
            &   \multicolumn{2}{c}{Predicted}  \\
    &       &   No &   Yes  & Total            \\ 
    \cline{2-5}
\multirow{2}{*}{\rotatebox[origin=c]{90}{Actual}}
    & No   & 7,202    & 656      & 7,858           \\
    & Yes  & 1,290    & 15,839    & 17,129          \\ 
    \cline{2-5}
    & Total & 8,492   & 16,495    & 24,987
\end{tabular}
\caption{Confusion matrix for prediction of anomalies for the matrix chain expression from benchmark data.}
\label{tbl::CF_MC4}
\end{table}

\subsection{The Expression $A A^T B$}\label{sec:result-other}

\subsubsection{Abundance}\label{sec:aatb_rand_anom}

\begin{figure}[tbh]
    \centering
    \includegraphics[width=\columnwidth]{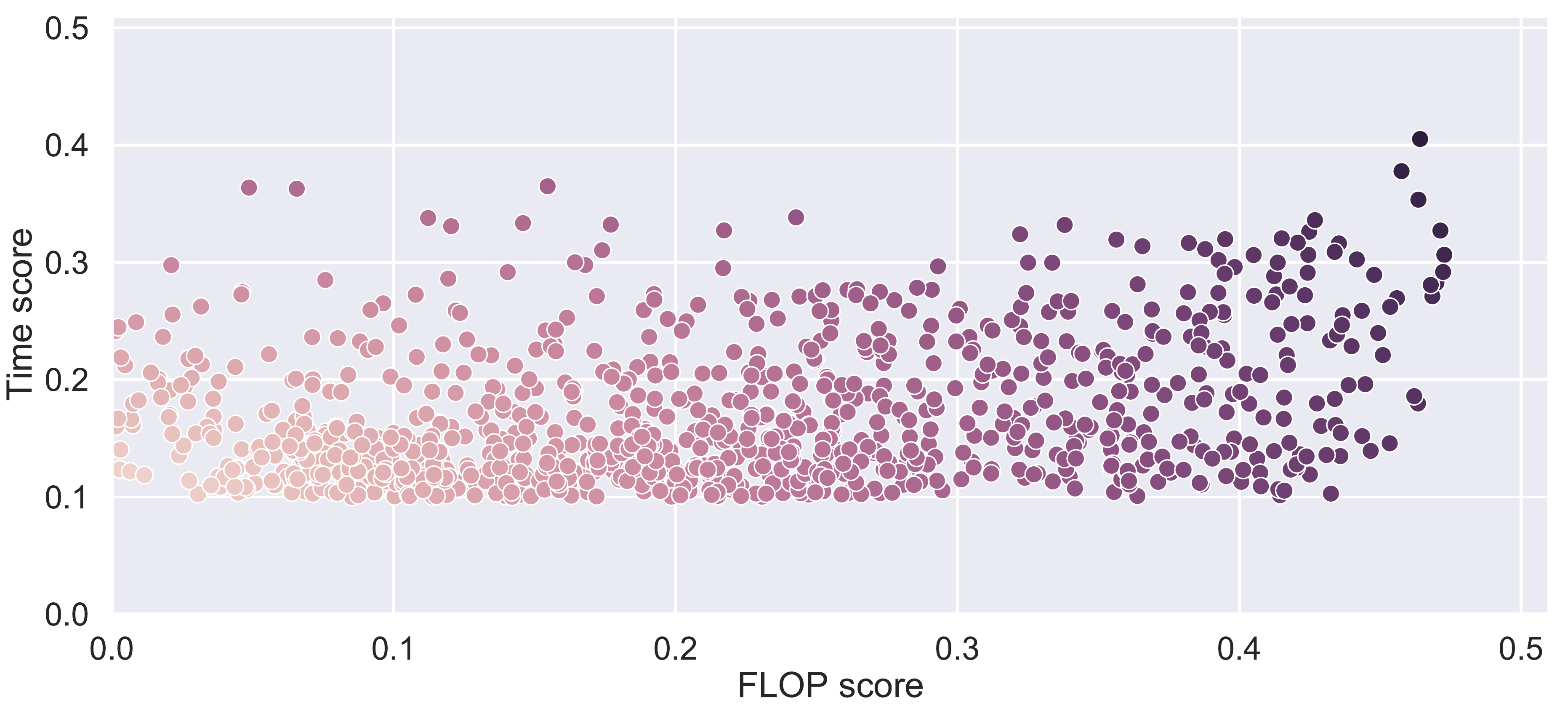}
    \caption{Scatter plot of the time score versus FLOP score for 1,000 anomalies of $A A^T B$.}
    \label{fig::aatb_randomAnomalies}
\end{figure}

The time score threshold for \hyperref[sec:exp-1]{Experiment~1} was set to $10\%$ (see Section~\ref{sec:anomaly-classification}) and the search continued until 1,000 anomalies were found. 
The search was restricted to the box $20 \leq d_i \leq 1200$ for $i = 0, 1, 2$.
Figure~\ref{fig::aatb_randomAnomalies} shows the results in the form of a scatter plot of time scores versus FLOP scores.
A total of 10,258 samples were required to find 1,000 anomalies, which translates to an abundance of $9.7\%$ in the constrained search space.
In total, $39.2\%$ of the anomalies present a time score above $20\%$ or a FLOP score above $30\%$.
In some extreme instances, performing $45\%$ more FLOPs reduces the execution time by $40\%$.

\subsubsection{Regions} 

\begin{figure}[tbh]
    \centering
    \includegraphics[width=\columnwidth]{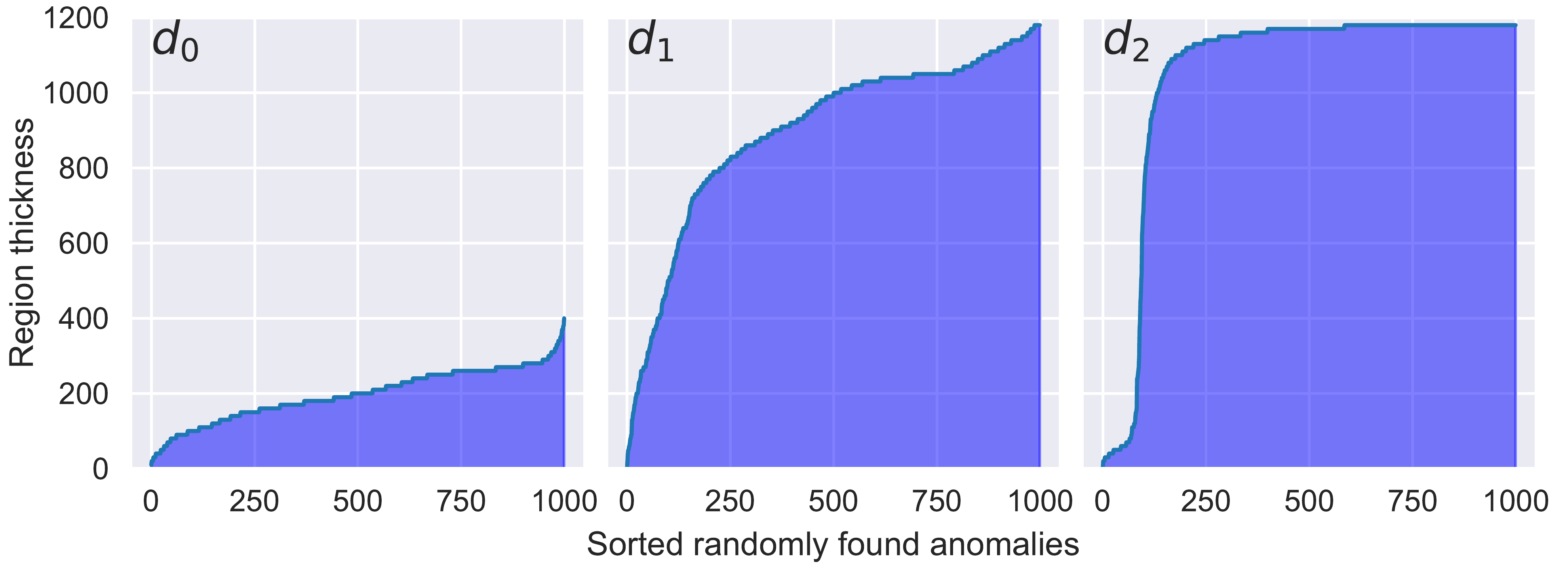}
    \caption{Distribution of the thickness of the regions around 1,000 anomalies of $A A^T B$ in each of the three dimensions \fix{from $d_0$ (left) to $d_2$ (right).}}
    \label{fig::aatb_Dim_Expl}
\end{figure}

Figure~\ref{fig::aatb_Dim_Expl} shows the distribution of the thickness of the regions in each of the three dimensions derived using data from \hyperref[sec:exp-2]{Experiment~2}.  
The regions are significantly thinner in dimension $d_0$ compared to the other dimensions.

\subsubsection{Region Boundaries}

\begin{figure}[tbh]
    \centering
    \includegraphics[width=\columnwidth]{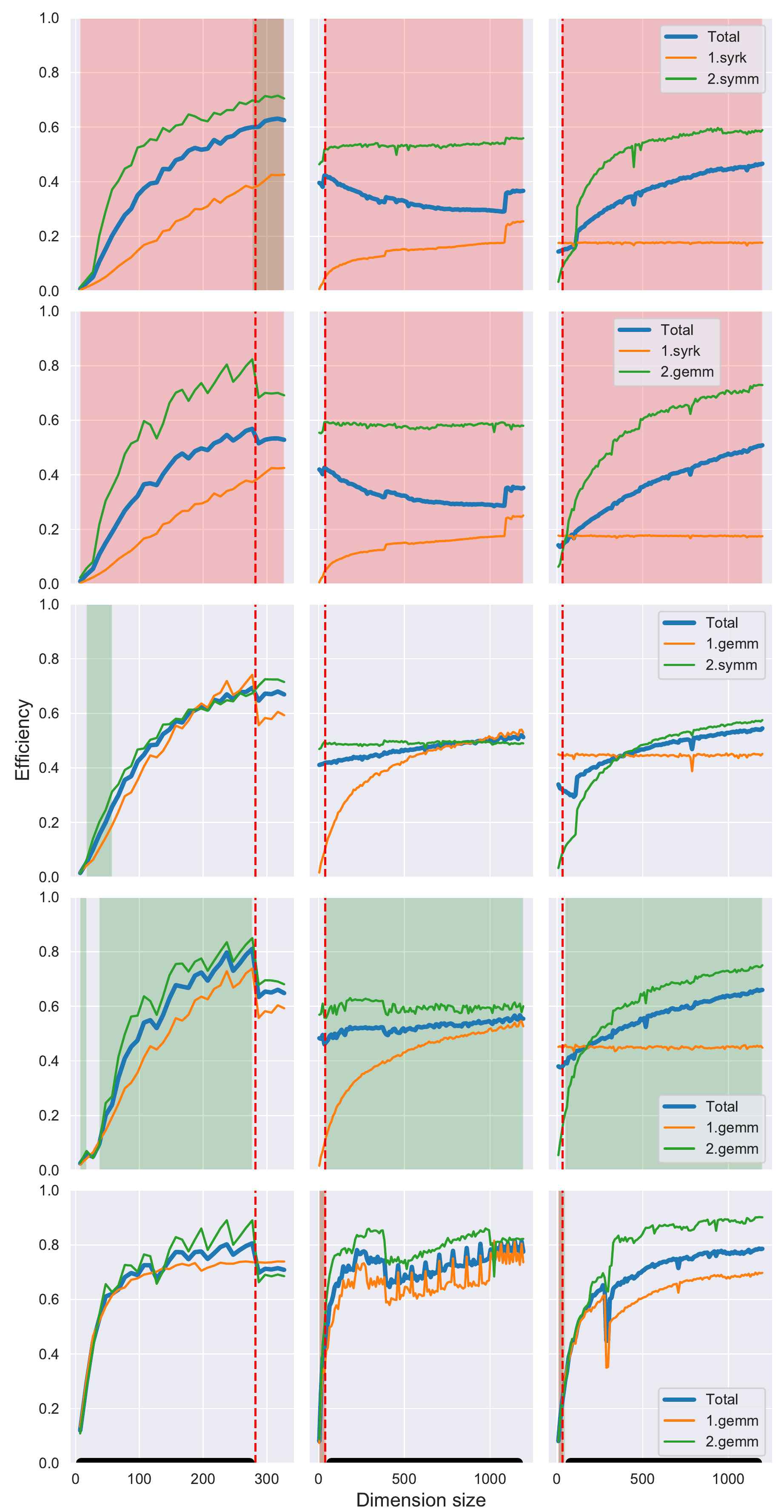}
    \caption{\fix{Three examples (left to right) of efficiencies in anomalous regions in the three dimensions of $A A^T B$.
    Each row corresponds to one of the five algorithms.
    Each column is a different anomaly and dimension traversed.
    Left: Efficiency along the line $(227 \pm 10 x,260,549)$, dimension $d_0$ being traversed.
    Center: Efficiency along the line $(80,514 \pm 10 x,768)$, dimension $d_1$ being traversed.
    Right: Efficiency along the line $(110,301,938 \pm 10 x)$, dimension $d_2$ being traversed.}
    }
    \label{fig::aatb_Eff_Dim}
\end{figure}

Figure~\ref{fig::aatb_Eff_Dim} uses data from \hyperref[sec:exp-2]{Experiment~2} to illustrate, for three different anomalies, how the efficiencies of the five algorithms change along one of the dimensions. 
Each column is a separate example with its own anomaly and traversed dimension.
\fix{The anomalies were picked at random.}
The rows from top to bottom correspond to Algorithms~1 through~5.  
Besides each algorithm's efficiency, the figure also plots the efficiency of each of the two kernel calls per algorithm.
Recall that the algorithms use different kernels (see Section~\ref{sec:aatb}).
The background colour convention is the same as in Figure~\ref{fig::mc4_eff_dim}.

Algorithms~1 and~2 have the same FLOP counts and are the cheapest throughout the anomalous regions in Figure~\ref{fig::aatb_Eff_Dim}. 
Algorithms~3 and~4 also have identical FLOP counts, but they are always strictly more expensive than Algorithms~1 and~2 (see Section~\ref{sec:aatb}). 

For the anomaly in the leftmost column, the line is in dimension $d_0$. 
The anomalous region covers $d_0 \leq 290$, and Algorithms~3 and~4 take turns being fastest while Algorithms~1 and~2 are cheapest. 
For $d_0 > 290$, Algorithm~1 is cheapest \emph{and} fastest and, therefore, there is no anomaly here. 

In the middle column, the line is in dimension $d_1$.
For values of $d_1$ close to zero, Algorithm~5 is both cheapest and fastest (barely visible), meaning there is no anomaly in this segment.
The region spans $d_1 \geq 40$ until the boundary of the search space.
Entering the region near $d_1 = 40$, Algorithms~1 and~2 become cheapest, but Algorithm~4 is the fastest throughout.

Finally, for the third anomaly in the rightmost column, the line is in dimension $d_2$.
The region covers the full segment except for tiny values of $d_2$. 
When $d_2$ is close to zero, Algorithm~5 is both cheapest and fastest (barely visible). 
For larger values of $d_2$, Algorithm~4 is fastest while Algorithms~1 and~2 remain cheapest.

Similar to the matrix chain expression, these examples also illustrate both types of transitions. 
In other words, some transitions are due to abrupt changes in efficiency while others are not.

\subsubsection{Prediction from Benchmarks}

Data from \hyperref[sec:exp-3]{Experiment~3} allows us to determine how well the anomalies discovered in \hyperref[sec:exp-2]{Experiment~2} could have been predicted from benchmarked performance profiles of the kernels used in the algorithms. 
The time score threshold was set at $5\%$. 
The result of the experiment is summarised in the form of a confusion matrix in Table~\ref{tbl::AATB_CF}.
For this expression, $75\%$ of the anomalies could have been predicted from the benchmark data alone, and $98.5\%$ of the predicted anomalies were actual anomalies. 

\begin{table}[htbp]
\begin{tabular}{cc|cc|c}
\multicolumn{2}{c}{}
            &   \multicolumn{2}{c}{Predicted}  \\
    &       &   No &   Yes  & Total            \\ 
    \cline{2-5}
\multirow{2}{*}{\rotatebox[origin=c]{90}{Actual}}
    & No   & 36,041    & 2,434      & 38,475           \\
    & Yes  & 53,711    & 160,867    & 214,578         \\ 
    \cline{2-5}
    & Total & 89,752   & 163,301    & 253,053
\end{tabular}
\caption{Confusion matrix for prediction of anomalies for the $A A^T B$ expression from benchmark data.}
\label{tbl::AATB_CF}
\end{table}

\section{Conclusions and Future Work}\label{sec:conclusion}

The process of mapping a linear algebra expression into a sequence of calls to numerical libraries such as BLAS and LAPACK can be seen as consisting of two steps. 
First, generate a set of mathematically equivalent algorithms, each of which evaluates the expression.
Second, given a concrete instance and computer system, select from this set one algorithm that is likely to be the fastest and use that to evaluate the expression.
In the simplest case where the sizes of all operands are known, then, at least in principle, the fastest algorithm can be selected through brute force search.
As soon as one or more of the sizes are symbolic at compile time, algorithm selection must be delayed until run time. 
Regardless if it is a matter of efficiency (fixed sizes) or necessity (symbolic sizes), there is value in procedures for algorithm selection that do not rely on empirical testing. 
Systems that address this problem, such as Transfor and Linnea, use FLOP counts as a discriminant.

In this paper, we experimentally studied how frequently and why FLOP count fails to be a reliable discriminant when mapping dense linear algebra expressions to the BLAS.
We ran a set of three experiments (see Section~\ref{sec:experiments}) on two expressions: The matrix chain expression $A B C D$ and the expression $A A^T B$.

We first estimated the abundance of anomalies through random search (\hyperref[sec:exp-1]{Experiment~1}).
Firstly, we searched for anomalies in the matrix chain expression, whose algorithms only involve the kernel that is regarded as the most highly optimised in BLAS (GEMM).
Even in this limiting case, we found that anomalies do exist although they appear to be rare (0.4\%).
Secondly, we did the same for the expression $A A^T B$. 
In comparison with the matrix chain expression, this one presents fewer algorithms, but those are built from a more varied set of BLAS kernels.
The results from both expressions exhibit a sheer contrast, since anomalies appeared to be much more frequent in the second expression (9.7\%).
Both expressions are very simple compared to those typically encountered in the wild. 
That such simple expressions can still have an abundance of anomalies leads us to expect that anomalies will be even more frequent in more complex expression.
This is motivated by the fact that large expressions have many more mathematically equivalent algorithms and also involve more kernels. 
These are two factors that one can reasonably assume will increase the opportunities for anomalies to occur.

% Regions.
In this study, we have also confirmed that anomalies tend to cluster together in contiguous regions, some of them extending throughout the entire range for one or more dimension sizes.
The boundaries of a region sometimes coincide with abrupt changes in the efficiency of one or more algorithms. 
Other times the onset of a region is due to more gradual changes in the kernels' efficiencies. 
We conjecture that knowledge of the location of abrupt changes in the performance profiles of the kernels will help to localise regions of severe anomalies. 
This information might be particularly useful to solve the LAMP with symbolic sizes.

% Performance profiles.
Finally, we leveraged the data generated when traversing the regions to investigate the underlying causes of the anomalies.
Although we know that an algorithm's execution time is determined by the interplay of its FLOP count, the performance of the invoked kernels, and the inter-kernel caching effects, we sought to determine their weights in the occurrence of anomalies.
In doing so, we quantified for both expressions how many of these anomalies could have been predicted solely from the benchmark data.
The high fraction of actual anomalies that could have been predicted from benchmark data (92\% and 75\%, for both expressions, respectively) leads us to conjecture that the kernels performance profiles have a greater weight on the anomalies.

In conclusion, this work has shown that FLOP count is not an adequate discriminant \fix{even when selecting amongst algorithms built from optimised libraries for simple expressions such as $A A^T B$}.
\fix{Although our experiments used a specific computer and implementation of BLAS, the qualitative conclusions are likely to generalise.
A different setup will affect the performance profiles of the  kernels, which, in turn, will translate into the disappearance of some anomalies and the surge of new ones.}

A natural next step is to combine FLOP counts with performance profiles of kernels to develop a methodology for localising regions of severe anomalies. 
A second step is to select algorithms based on more than the FLOP count; in particular, including performance profiles of kernels.  
Combined, these steps may lead to a more robust algorithm selection methodology suitable for complex expressions or expressions with symbolic sizes. 
Given that accurate performance modelling for sequences of kernels is known to be expensive and difficult~\cite{peise2014study}, finding ways to reduce the cost and complexity are expected to be crucial steps towards a practical algorithm selection methodology. 
In doing so, we will take another step towards selecting optimal algorithms to evaluate linear algebra expressions, and will be closer to squeeze the most out of our computational capabilities.

\bibliographystyle{ACM-Reference-Format}
\bibliography{Source/bibliography.bib}

\end{document}